\documentclass{IEEEcsmag}

\usepackage[colorlinks,urlcolor=blue,linkcolor=blue,citecolor=blue]{hyperref}
\expandafter\def\expandafter\UrlBreaks\expandafter{\UrlBreaks\do\/\do\*\do\-\do\~\do\'\do\"\do\-}
\usepackage{upmath}
\usepackage[dvipsnames]{xcolor}

\newcommand{\statbox}[2]{\noindent\raisebox{0.7\height}{\fcolorbox[rgb]{0,0,0}{#1}{\scriptsize~}}}

\newcommand{\circled}[1]{\raisebox{.4pt}{\textcircled{\raisebox{0.15pt}{\resizebox{!}{0.8ex}{\textbf{\textsf{#1}}}}}}}

\usepackage{longfbox}

\newfboxstyle{tight2}{padding=2pt,margin=0pt,baseline-skip=false}%

\definecolor{ColT1}{HTML}{1f77b4}
\definecolor{ColT2}{HTML}{ff7f0e}
\definecolor{ColT3}{HTML}{2ca02c}
\definecolor{ColT4}{HTML}{d62728}
\definecolor{ColT5}{HTML}{9467bd}
\definecolor{ColT6}{HTML}{8c564b}
\definecolor{ColT7}{HTML}{e377c2}
\definecolor{ColT8}{HTML}{7f7f7f}
\definecolor{ColT9}{HTML}{bcbd22}
\definecolor{ColT10}{HTML}{17becf}

\newlength{\boxh}
\usepackage[absolute,overlay]{textpos}
\usepackage{pbox}
\usepackage{everypage}

\newcommand{\stamp}[1][© 2024 IEEE. This is the author's version of the article that has been published in IEEE Computer Graphics and Applications. The final version of this record is available at: \href{https://doi.org/10.1109/MCG.2024.3360881}{\color{blue}10.1109/MCG.2024.3360881}]{%
\begin{textblock*}{165mm}(50mm,30mm)
\centering%
\small%
\emph{#1}%
\end{textblock*}%
}

\AddEverypageHook{
	\ifnum\value{page}=14
	  \relax
	\else
	  \ifnum\value{page}=15
	   \relax
	  \else
	   	\stamp
	  \fi
	\fi
}

\let\clearpage\relax

\jvol{XX}
\jnum{XX}
\paper{8}
\jmonth{October}
\jname{IEEE Computer Graphics and Applications}
\jtitle{IEEE Computer Graphics and Applications}
\pubyear{2023}

\begin{document}

\sptitle{FEATURE ARTICLE}

\title{Visualization for Trust in Machine Learning Revisited: The State of the Field in 2023}

\author{Angelos Chatzimparmpas}
\affil{{}Northwestern University, Evanston, IL, 60208, USA}

\author{Kostiantyn Kucher}
\affil{{}Linköping University, Norrköping, 60233, Sweden}

\author{Andreas Kerren}
\affil{{}Linköping University, Norrköping, 60233, Sweden; Linnaeus University, Växjö, 35195, Sweden}

\markboth{FEATURE ARTICLE}{FEATURE ARTICLE}

\begin{abstract}\looseness-1Visualization for explainable and trustworthy machine learning remains one of the most important and heavily researched fields within information visualization and visual analytics with various application domains, such as medicine, finance, and bioinformatics. After our 2020 state-of-the-art report comprising 200 techniques, we have persistently collected peer-reviewed articles describing visualization techniques, categorized them based on the previously established categorization schema consisting of 119 categories, and provided the resulting collection of 542 techniques in an online survey browser. In this survey article, we present the updated findings of new analyses of this dataset as of fall 2023 and discuss trends, insights, and eight open challenges for using visualizations in machine learning. Our results corroborate the rapidly growing trend of visualization techniques for increasing trust in machine learning models in the past three years, with visualization found to help improve popular model explainability methods and check new deep learning architectures, for instance.
\end{abstract}

\stamp

\maketitle

\chapteri{T}rust in machine learning (ML) models is a major concern in leveraging these technologies for real-world applications.$^1$ Yet, ML models are being deployed in different application fields, and their role in decision-making processes is growing rapidly. Fields such as healthcare and criminal justice increasingly depend on ML models to make irreversible decisions that impact human lives.$^2$ However, the black-box nature of some ML models poses a threat to their adoption. Domain experts often hesitate to rely on ML models for high-risk decision-making, as the inability to understand their inner workings fosters mistrust.$^3$

In response to the outlined challenges, researchers in academia and industry have designed several innovative solutions. For example, Google's Explainable Artificial Intelligence (AI) Cloud and Descriptive mAchine Learning EXplanations (DALEX) package aims to improve the collaboration among domain experts to address the challenges posed by the complexity of AI. Except for all the visualization techniques analyzed in this survey article, recent frameworks set a foundation for developing techniques that facilitate users in communicating and externalizing their trust explicitly across varied ML stages and in understanding the complexities of human and AI interactions.$^4$$^,$$^5$ Other works bridge the significant gap between ML outputs and human cognition by promoting a cross-disciplinary approach and building a robust model designed to examine the sender’s explanation intention and its ensuing impact on the receiver's perception.$^6$$^,$$^7$

We base this work upon the findings of our previously published survey articles$^8$$^,$$^9$ and others that have stressed the need for visual analytics (VA) to improve trust and transparency in areas such as dimensionality reduction (DR),$^1$$^1$ deep learning (DL),$^1$$^2$$^,$$^1$$^3$ and ML in general.$^1$$^4$$^,$$^1$$^5$ After our 2020 state-of-the-art report (STAR) comprising 200 techniques,$^9$ we have been collecting peer-reviewed articles describing visualization techniques for enhancing trust in ML, categorizing them based on the previously established categorization schema of 18 groups and 119 categories in total, and providing the hand-curated compilation of 542 visualization techniques in an online survey browser available at:

\begin{center}
\url{https://trustmlvis.lnu.se}
\end{center}

Although this survey article follows the same styling and overall structure as our STAR published in 2020,$^9$ here, we made a new analysis of the updated dataset and discuss the most recent findings that provide us with an overview of the trust in ML visualization field as of fall 2023, including insights about the visualization community. The contributions of this article are:

\begin{itemize}
    \item the 542 techniques from the past 15 years categorized in \emph{trust levels (TLs)} of interactive ML (compared to 200 in the 2020 STAR);
    \item the trends and category correlations found using topic, temporal, correlation, and pattern mining (new compared to the 2020 STAR) analyses;
    \item the interactive survey browser that aids domain experts, ML experts, visualization researchers, and others in exploring the field's literature; and 
    \item the eight open challenges in VA for increasing the trustworthiness of the ML process (new compared to the six challenges of the 2020 STAR).
\end{itemize}

\begin{figure}
\centerline{\includegraphics[width=18.5pc]{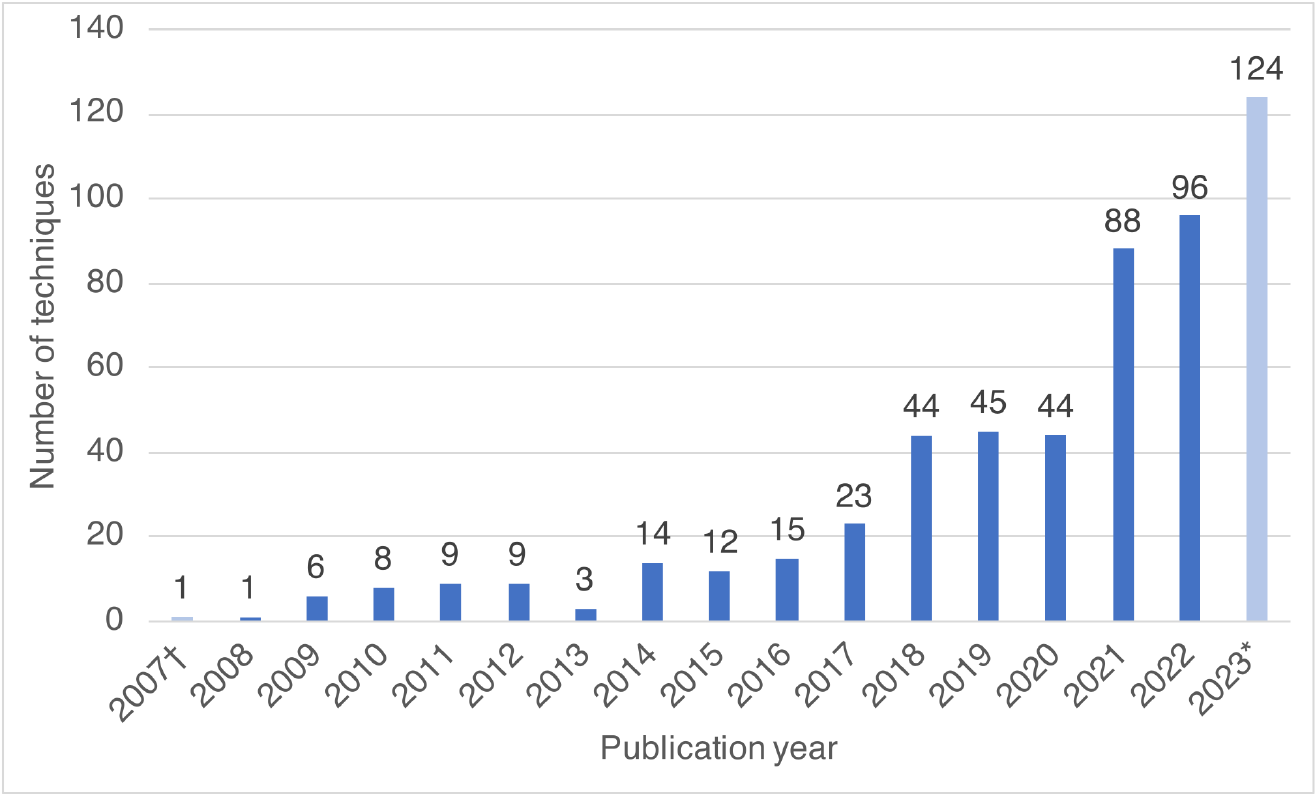}}
\caption{The 542 collected techniques by publication year.
\vspace{-2mm} \newline \scriptsize \emph{Note:}
\newline ($\ast$) The data collection was completed in September 2023. 
\newline ($\dagger$) This technique was found in one article's \emph{Related Work} section.}
\label{fig:time}
\end{figure}

\begin{table}
\caption{Number of visualization techniques \# by publication venues in either visualization (left) or other disciplines (right).}
\centerline{\includegraphics[width=18.5pc]{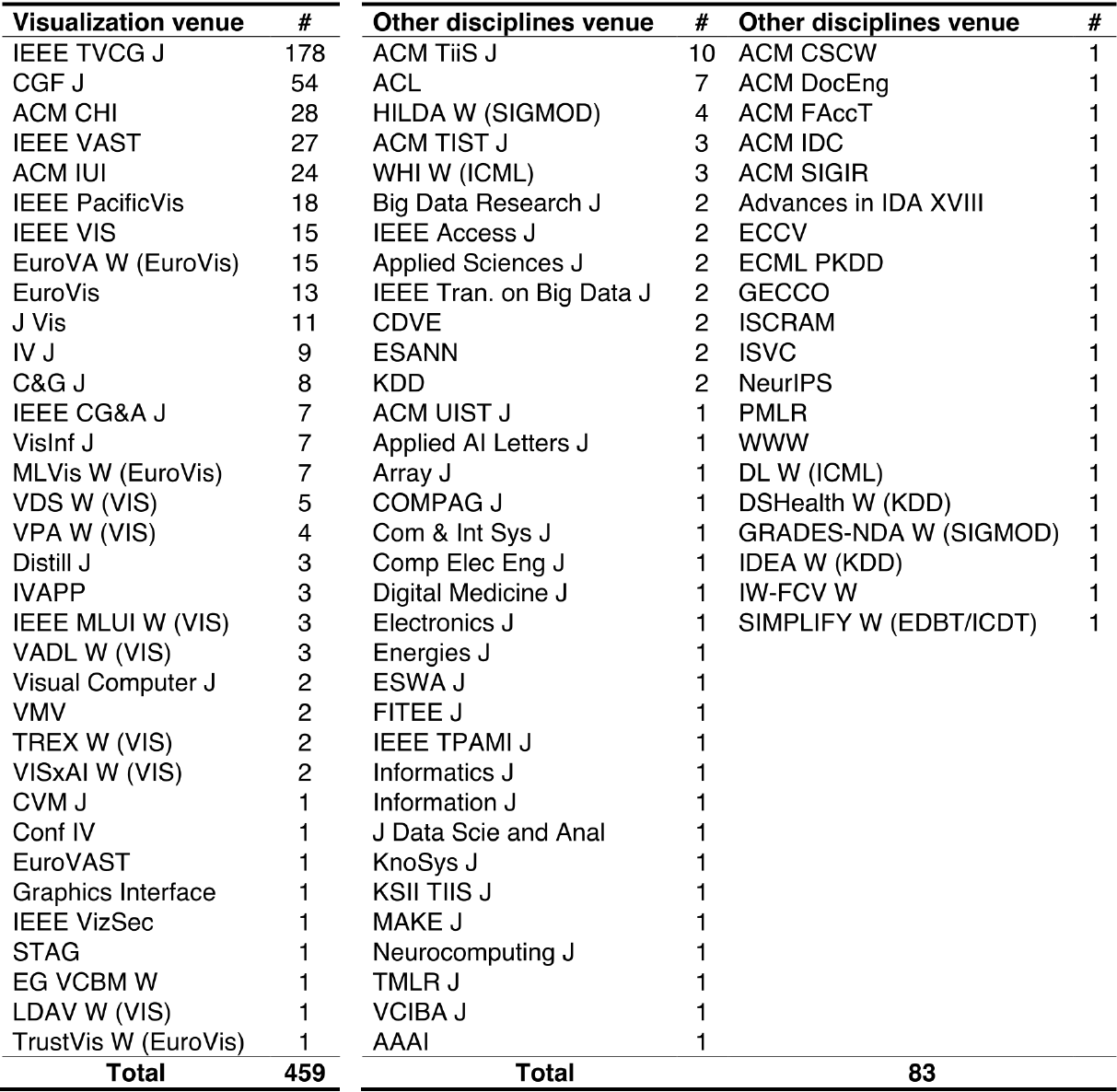}}
\smallskip \scriptsize \emph{Note:} Rows are conferences except if journals (`J') or workshops (`W').
\label{tab:venue} \vspace{-1.5mm}
\end{table}

\noindent In this survey article, we rely on the same literature search methodology and run an identical analysis as in our 2020 STAR.$^9$ We also adopt the same general definition of \emph{trust} by Lee and See:$^1$$^0$ ``the attitude that an agent will help achieve an individual’s goals in a situation characterized by uncertainty and vulnerability'' towards a more detailed, \emph{multi-level} model of trust in ML that consists of \emph{five trust levels}: \emph{the raw data} (TL1: \emph{source reliability} and \emph{transparent collection process}), \emph{the processed data} (TL2: \emph{uncertainty awareness}, \emph{equality/data bias}, \emph{comparison of structures}, \emph{guidance/re\-commendation}, and \emph{outlier detection}), \emph{the algorithm/learning method} (TL3: \emph{familiarity}, \emph{understanding/ explanation}, \emph{debugging/diagnosis}, \emph{refinement/steering}, \emph{comparison}, \emph{knowledgeability}, and \emph{fairness}), \emph{the concrete model(s) for a particular task} (TL4: \emph{experience}, \emph{in situ comparison}, \emph{performance}, \emph{what-if hypotheses}, \emph{model bias}, and \emph{model variance}), and \emph{the evaluation and the subjective users' expectations} (TL5: \emph{agreement of colleagues}, \emph{visualization evaluation}, \emph{metrics validation/results}, and \emph{user bias}). More details on methodology, trust levels, and our categorization can be found in our 2020 STAR.$^9$

\section{GENERAL OVERVIEW}

\subsection{Time \& Venues}
Our collection includes 542 techniques sourced from various journals, conferences, and workshops. Figure~\ref{fig:time} shows the temporal distribution of these publications, with a stable growth in interest in the topic since 2009 and a remarkable increase evident in 2021, as well as another significant rise in numbers for 2023.

Table~\ref{tab:venue} outlines the distribution of publication venues. The majority of the techniques were primarily published on visualization venues. Notable exceptions include workshops such as WHI of ICML and special journal issues such as the human-centered explainable AI of ACM TiiS, aiming to engage with the ML community. Despite these efforts, the limited number of publications in other discipline venues may imply that visualization researchers struggle with external outreach. This situation may also suggest that experts from other fields remain largely unaware of the extensive opportunities offered by the visualization field and highlight the importance of visualization literacy.

\begin{figure*}
\centerline{\includegraphics[width=37pc]{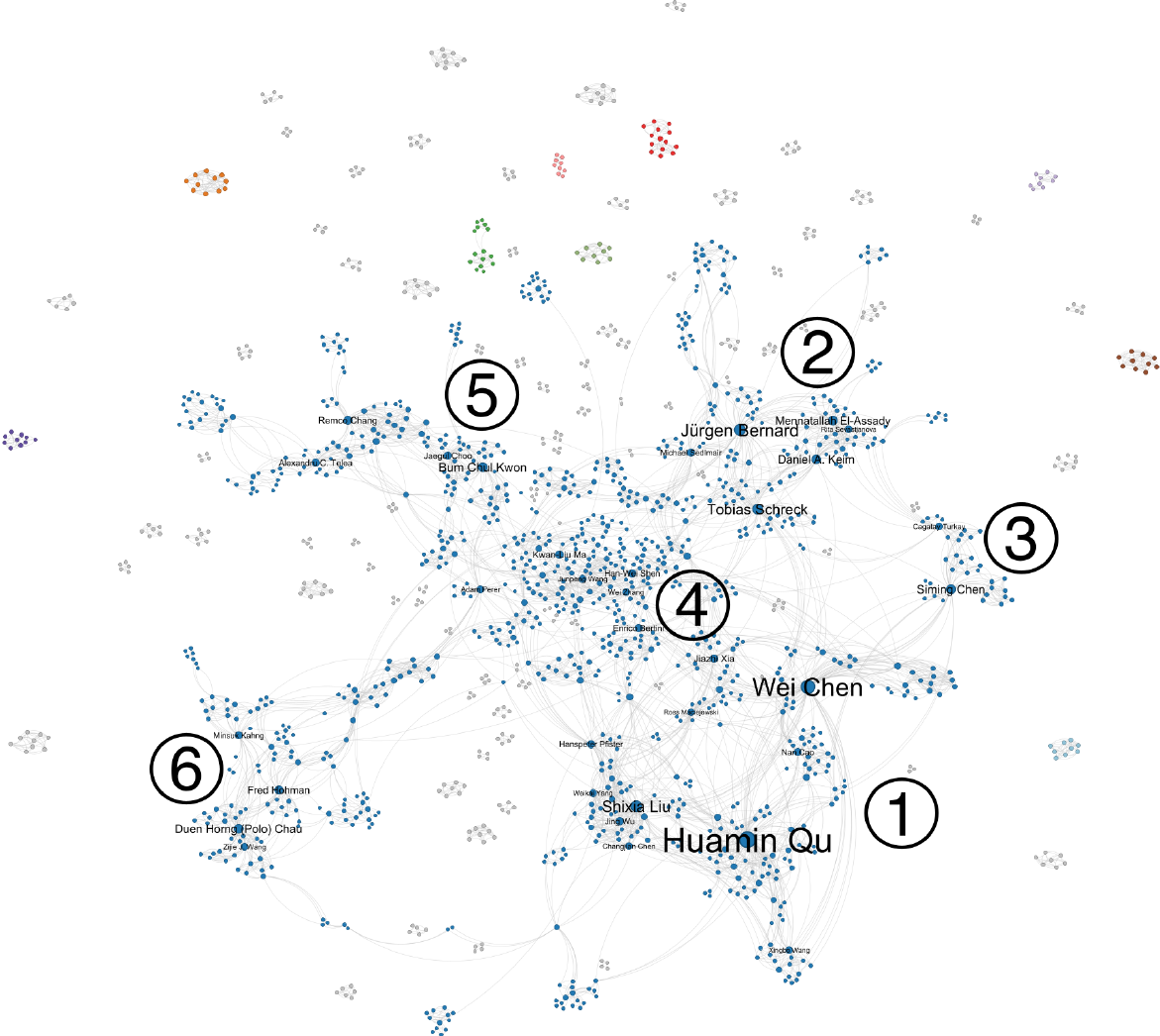}} \vspace{-2mm}
\caption{Co-authorship network for all articles, with the ten largest clusters in different colors. \circled{1}--\circled{6} are interesting subclusters within the blue cluster. The node size shows the in-degree centrality of each author, with the labels filtered to improve readability.} 
\label{fig:co-author} \vspace{-3mm}
\end{figure*}

\begin{table*}
\caption{Overview of datasets ordered according to their usage with more than four occurrences.}
\label{tab:data}
\centerline{\includegraphics[width=37pc]{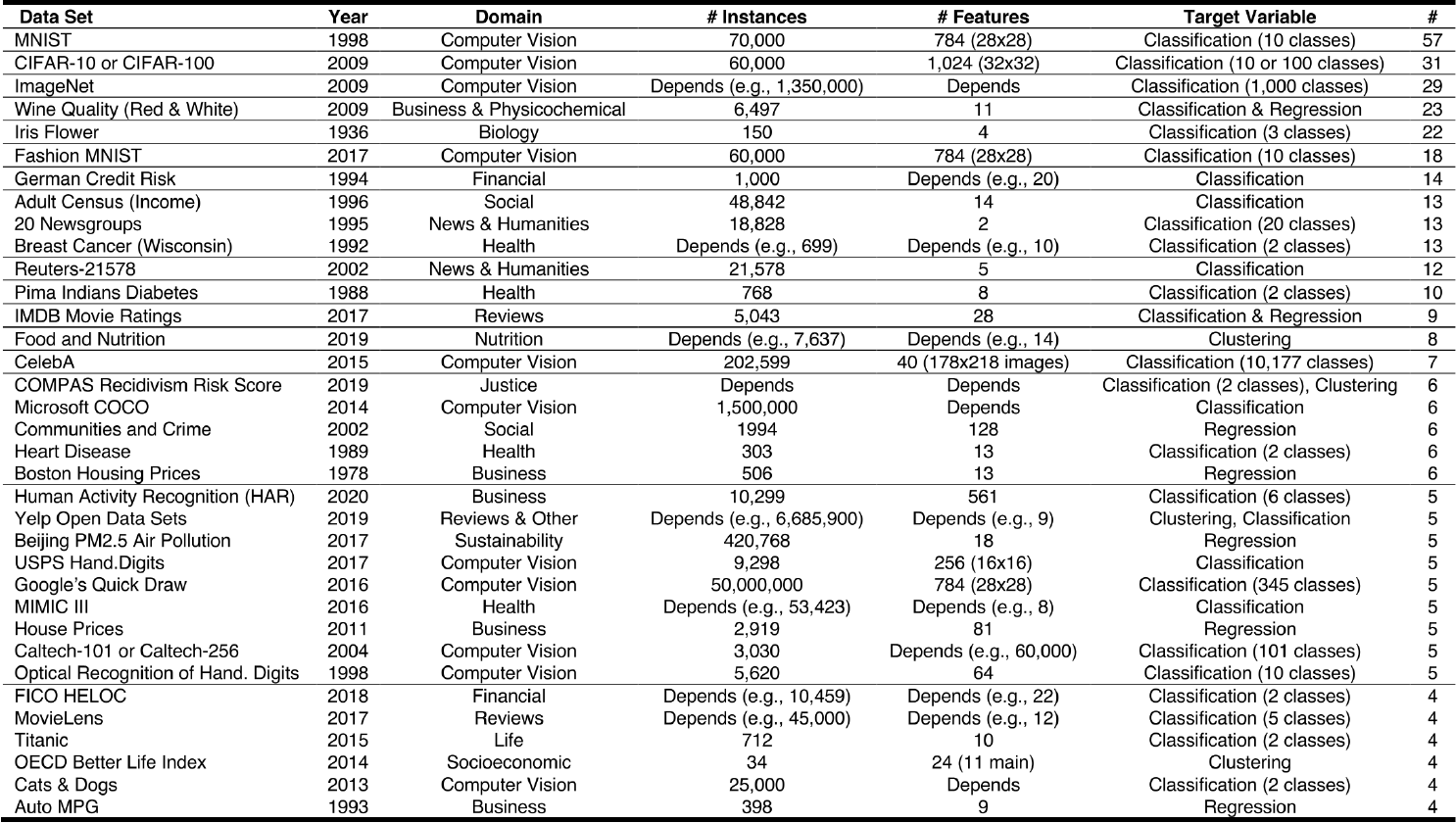}}
\medskip
\scriptsize \emph{Note:} `\#' column shows the number of articles in this survey using a specific dataset; the datasets are also grouped based on this frequency.
\end{table*}

\subsection{Co-authorship Network}
We performed co-authorship analysis in Gephi, weighting author nodes by their number of co-authorship edges (Figure~\ref{fig:co-author}). The network has 96 weakly connected components, with the ten largest color-coded.

A dominant blue cluster accounts for $\approx$$70\%$ of all network connections compared to $\approx$$41\%$ in the 2020 STAR. Within this cluster, the most influential authors in \circled{1} are all based in China: Huamin Qu (HKUST), Wei Chen (ZJU), and Shixia Liu (THU). \circled{2} features Jürgen Bernard (UZH, Switzerland), Tobias Schreck (TU Graz, Austria), Daniel A. Keim (U of Konstanz, Germany), and Mennatallah El-Assady (ETH Zurich, Switzerland) as the key names, with Siming Chen (Fudan U, China) and others as an in-between node shown in \circled{3}. Prominent authors in \circled{4} include Kwan-Liu Ma (UC Davis, USA), Jiazhi Xia (CSU, China), Han-Wei Shen (OSU, USA), and Enrico Bertini (NEU, USA). \circled{5} is relatively isolated compared to \circled{4} and comprises Bum Chul Kwon (IBM Research, USA), Remco Chang (Tufts U, USA), Jaegul Choo (KAIST, South Korea), and Alexandru C. Telea (UU, The Netherlands). \circled{6} includes Fred Hohman (Apple Research), Minsuk Kahng (Google Research), and Bongshin Lee (Microsoft Research) from the industry based in the USA, along with academic collaborators like Duen Horng (Polo) Chau and Zijie J. Wang from Georgia Tech, USA. This industry subcluster stands somewhat apart from its academic counterparts. Although it is more connected than in the 2020 STAR, this finding suggests that increased collaboration between these two sectors may improve the visualization research output. The network also contains many smaller, color-coded clusters of collaborative researchers (mostly from the same labs). An analysis of authorship count distributions shows $\approx$$75\%$ unique authors out of 602 for 2020 and $\approx$$73\%$ of 1,539 authors for 2023, indicating a similar trend for the overall authorship (see supplemental material).

\subsection{Datasets}
We analyzed publicly accessible datasets from the 542 articles, sorted by frequency and then by recency. Table~\ref{tab:data} lists 35 datasets, found in at least four articles.

The most frequently occurring datasets are MNIST, CIFAR-10/100, ImageNet, Wine Quality, Iris Flower, Fashion MNIST, German Credit Risk, Adult Census, 20 Newsgroups, and Breast Cancer. Of these ten datasets, four are about computer vision and are typically used in articles related to DL and neural networks (NNs). \emph{Classification} is by far the most frequent \emph{target variable}, followed by \emph{regression} and \emph{clustering}. Table~\ref{tab:data} contains information about the number of instances and features, as well as the number of classes (if there are any available). The number of articles that used the datasets is visible in the `\#' column of Table~\ref{tab:data}.

\subsection{Survey Browser}
\begin{figure*}
\centerline{\includegraphics[width=37pc]{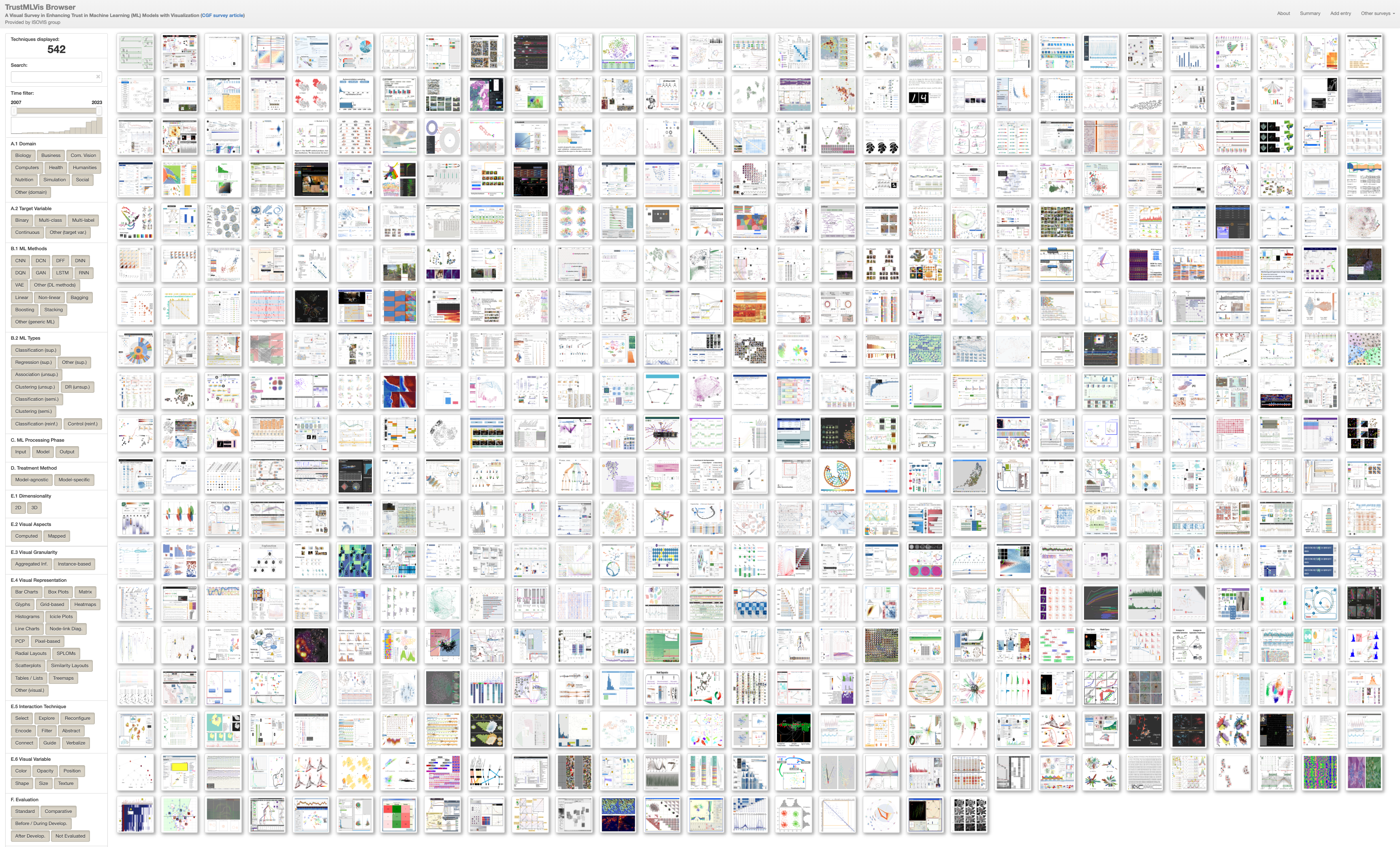}} \vspace{-2mm}
\caption{Our interactive survey browser for easily exploring all identified visualization techniques (available at \href{https://trustmlvis.lnu.se}{trustmlvis.lnu.se}).}
\label{fig:web} \vspace{-1mm}
\end{figure*}

The interactive survey browser's user interface (UI) consists of (1) a grid showcasing thumbnails of various visualization techniques and (2) an interactive panel that facilitates users to filter based on categories, time, and text (see Figure~\ref{fig:web}). Clicking a thumbnail reveals details and bibliographic references for that particular technique. At the top of the webpage, links provide access to dialogs containing detailed statistics for the dataset and supplemental materials. We encourage readers to explore the online survey browser and suggest new articles via the "Add entry" dialog.

\begin{table*}[t!]
\centering
	\caption{ML visualization techniques categorization with the 2023 data (including comparison with the 2020 STAR data).}
	\setlength{\tabcolsep}{2pt}
	\renewcommand{\arraystretch}{1}
	\setlength\doublerulesep{2mm} 
	\tiny
	\begin{minipage}[t]{0.36\textwidth}
	\begin{tabular}[t]{|l|r|c|}
	\hline \textbf{Domain} & \textbf{542}~\statbox{0.00,0.50,0.00}{100\%} & \\
	\hline Biology & 49~\statbox{0.91,0.96,0.91}{9\%} & -5\%~$\downarrow$\\ 
	\hline Business & 87~\statbox{0.84,0.92,0.84}{16\%} & +6\%~$\uparrow$\\
	\hline Computer Vision & 180~\statbox{0.67,0.84,0.67}{33\%} & +3\%~$\uparrow$\\
	\hline Computers & 9~\statbox{0.98,0.99,0.98}{2\%} & -1\%~$\downarrow$\\
	\hline Health & 90~\statbox{0.84,0.92,0.84}{17\%} & +2\%~$\uparrow$\\
	\hline Humanities & 81~\statbox{0.85,0.93,0.85}{15\%} & -6\%~$\downarrow$\\
	\hline Nutrition & 16~\statbox{0.97,0.98,0.97}{3\%} & -1\%~$\downarrow$\\
	\hline Simulation & 19~\statbox{0.96,0.98,0.96}{4\%} & 0\%~~-\\
	\hline Social / Socioeconomic & 58~\statbox{0.89,0.95,0.89}{11\%} & 0\%~~-\\
	\hline Other & 230~\statbox{0.58,0.79,0.58}{42\%} & -5\%~$\downarrow$\\
	\hline 
	\hline \textbf{Target Variable} & \textbf{542}~\statbox{0.00,0.50,0.00}{100\%} & \\ 
	\hline Binary (categorical) & 85~\statbox{0.84,0.92,0.84}{16\%} & -4\%~$\downarrow$\\
	\hline Multi-class (categorical) & 297~\statbox{0.45,0.73,0.45}{55\%} & -9\%~$\downarrow\downarrow$\\
	\hline Multi-label (categorical) & 21~\statbox{0.96,0.98,0.96}{4\%} & -1\%~$\downarrow$\\
	\hline Continuous (regression problems) & 60~\statbox{0.88,0.94,0.88}{12\%} & 0\%~~-\\
	\hline Other & 147~\statbox{0.73,0.87,0.73}{27\%} & +8\%~$\uparrow$\\
	\hline 
	\hline \textbf{ML Methods} & \textbf{542}~\statbox{0.00,0.50,0.00}{100\%} & \\ 
	\hline Convolutional Neural Network (CNN) & 72~\statbox{0.87,0.93,0.87}{13\%} & 0\%~~-\\
	\hline Deep Convolutional Network (DCN) & 12~\statbox{0.98,0.99,0.98}{2\%} & -2\%~$\downarrow$\\
	\hline Deep Feed Forward (DFF) & 11~\statbox{0.98,0.99,0.98}{2\%} & -3\%~$\downarrow$\\
	\hline Deep Neural Network (DNN) & 55~\statbox{0.90,0.95,0.90}{10\%} & 0\%~~-\\
	\hline Deep Q-Network (DQN)& 11~\statbox{0.98,0.99,0.98}{2\%} & -3\%~$\downarrow$\\
	\hline Generative Adversarial Network (GAN) & 13~\statbox{0.98,0.99,0.98}{2\%} & -3\%~$\downarrow$\\
	\hline Long Short-Term Memory (LSTM) & 26~\statbox{0.95,0.98,0.95}{5\%} & -2\%~$\downarrow$\\
	\hline Recurrent Neural Network (RNN) & 31~\statbox{0.94,0.97,0.94}{6\%} & -3\%~$\downarrow$\\
	\hline Variational Auto-Encoder (VAE) & 22~\statbox{0.96,0.98,0.96}{4\%} & -3\%~$\downarrow$\\
	\hline Other (DL methods) & 92~\statbox{0.84,0.92,0.84}{17\%} & +6\%~$\uparrow$\\
	\hline Linear (DR) & 81~\statbox{0.85,0.93,0.85}{15\%} & -14\%~$\downarrow\downarrow$\\
	\hline Non-linear (DR) & 94~\statbox{0.83,0.91,0.83}{17\%} & -9\%~$\downarrow\downarrow$\\
	\hline Bagging (ensemble learning) & 74~\statbox{0.86,0.93,0.86}{14\%} & 0\%~~-\\
	\hline Boosting (ensemble learning) & 30~\statbox{0.94,0.97,0.94}{6\%} & 0\%~~-\\
	\hline Stacking (ensemble learning) & 10~\statbox{0.98,0.99,0.98}{2\%} & -1\%~$\downarrow$\\
	\hline Other (generic)& 233~\statbox{0.57,0.78,0.57}{43\%} & -6\%~$\downarrow$\\
	\hline 
	\hline \textbf{ML Types} & \textbf{500}~\statbox{0.08,0.54,0.08}{92\%} & \\ 
	\hline Classification (supervised)& 288~\statbox{0.47,0.74,0.47}{53\%} & -3\%~$\downarrow$\\
	\hline Regression (supervised) & 55~\statbox{0.90,0.95,0.90}{10\%} & 0\%~~-\\
	\hline Other (supervised)& 32~\statbox{0.95,0.97,0.95}{6\%} & +2\%~$\uparrow$\\
	\hline Association (unsupervised) & 14~\statbox{0.97,0.99,0.97}{3\%} & 0\%~~-\\
	\hline Clustering (unsupervised)& 69~\statbox{0.87,0.94,0.87}{13\%} & -8\%~$\downarrow$\\
	\hline Dimensionality Reduction (unsupervised)& 104~\statbox{0.81,0.90,0.81}{19\%} & -14\%~$\downarrow\downarrow$\\
	\hline Classification (semi-supervised)& 40~\statbox{0.93,0.96,0.93}{7\%} & 0\%~~-\\
	\hline Clustering (semi-supervised)& 16~\statbox{0.97,0.98,0.97}{3\%} & 0\%~~-\\
	\hline Classification (reinforcement)& 4~\statbox{1.00,1.00,1.00}{1\%} & 0\%~~-\\
	\hline Control (reinforcement)& 11~\statbox{0.98,0.99,0.98}{2\%} & 0\%~~-\\
	\hline	
	\end{tabular} 
	\bigbreak
	Color Legend: \thinspace
	\statbox{1.00,1.00,1.00}{0\%} {\strut 0 articles} \thinspace
	\statbox{0.50,0.75,0.50}{50\%} {\strut 271 articles} \thinspace
	\statbox{0.00,0.50,0.00}{100\%} {\strut 542 articles}
	\end{minipage}
	\hspace{1mm}
	\begin{minipage}[t]{0.33\textwidth}
	\begin{tabular}[t]{|l|r|c|}
	\hline \textbf{ML Processing Phase} & \textbf{542}~\statbox{0.00,0.50,0.00}{100\%} & \\ 
	\hline Pre-processing / Input & 120~\statbox{0.78,0.89,0.78}{22\%} & +4\%~$\uparrow$\\
	\hline In-processing / Model & 156~\statbox{0.71,0.85,0.71}{29\%} & +6\%~$\uparrow$\\
	\hline Post-processing / Output & 363~\statbox{0.33,0.67,0.33}{67\%} & -14\%~$\downarrow\downarrow$\\
	\hline
	\hline \textbf{Treatment Method} & \textbf{542}~\statbox{0.00,0.50,0.00}{100\%} & \\ 
	\hline Model-agnostic / Black Box & 395~\statbox{0.27,0.64,0.27}{73\%} & +1\%~$\uparrow$\\
	\hline Model-specific / White Box & 165~\statbox{0.69,0.85,0.69}{30\%} & -5\%~$\downarrow$\\
	\hline 
	\hline \textbf{Dimensionality} & \textbf{542}~\statbox{0.00,0.50,0.00}{100\%} & \\ 
	\hline 2D & 538~\statbox{0.01,0.51,0.01}{99\%} & +1\%~$\uparrow$\\
	\hline 3D & 9~\statbox{0.98,0.99,0.98}{2\%} & -1\%~$\downarrow$\\
	\hline
	\hline \textbf{Visual Aspects} & \textbf{542}~\statbox{0.00,0.50,0.00}{100\%} & \\ 
	\hline Computed & 497~\statbox{0.08,0.55,0.08}{92\%} & -6\%~$\downarrow$\\
	\hline Mapped & 273~\statbox{0.50,0.75,0.50}{50\%} & -5\%~$\downarrow$\\
	\hline 
	\hline \textbf{Visual Granularity} & \textbf{542}~\statbox{0.00,0.50,0.00}{100\%} & \\ 
	\hline Aggregated Information & 468~\statbox{0.14,0.57,0.14}{86\%} & -6\%~$\downarrow$\\
	\hline Instance-based / Individual & 377~\statbox{0.31,0.65,0.31}{70\%} & -3\%~$\downarrow$\\
	\hline 
	\hline \textbf{Visual Representation} & \textbf{542}~\statbox{0.00,0.50,0.00}{100\%} & \\ 
	\hline Bar Charts & 243~\statbox{0.55,0.77,0.55}{45\%} & +4\%~$\uparrow$\\
	\hline Box Plots & 39~\statbox{0.93,0.96,0.93}{7\%} & +1\%~$\uparrow$\\
	\hline Matrix & 153~\statbox{0.72,0.86,0.72}{28\%} & +3\%~$\uparrow$\\
	\hline Glyphs / Icons / Thumbnails & 131~\statbox{0.76,0.88,0.76}{24\%} & -8\%~$\downarrow$\\
	\hline Grid-based Approaches & 59~\statbox{0.89,0.95,0.89}{11\%} & +1\%~$\uparrow$\\
	\hline Heatmaps & 149~\statbox{0.73,0.86,0.73}{27\%} & +4\%~$\uparrow$\\
	\hline Histograms & 174~\statbox{0.68,0.84,0.68}{32\%} & +4\%~$\uparrow$\\
	\hline Icicle Plots & 9~\statbox{0.98,0.99,0.98}{2\%} & -1\%~$\downarrow$\\
	\hline Line Charts & 154~\statbox{0.72,0.86,0.72}{28\%} & 0\%~~-\\
	\hline Node-link Diagrams & 128~\statbox{0.76,0.88,0.76}{24\%} & 0\%~~-\\
	\hline Parallel Coordinates Plots (PCPs) & 73~\statbox{0.87,0.93,0.87}{13\%} & -3\%~$\downarrow$\\
	\hline Pixel-based Approaches & 20~\statbox{0.96,0.98,0.96}{4\%} & 0\%~~-\\
	\hline Radial Layouts & 90~\statbox{0.84,0.92,0.84}{17\%} & +6\%~$\uparrow$\\
	\hline Scatterplot Matrices (SPLOMs) & 26~\statbox{0.95,0.97,0.95}{5\%} & -4\%~$\downarrow$\\
	\hline Scatterplot / Projections & 294~\statbox{0.46,0.73,0.46}{54\%} & -4\%~$\downarrow$\\
	\hline Similarity Layouts & 190~\statbox{0.65,0.82,0.65}{35\%} & +21\%~$\uparrow\uparrow\uparrow$\\
	\hline Tables / Lists & 201~\statbox{0.63,0.82,0.63}{37\%} & -6\%~$\downarrow$\\
	\hline Treemaps & 13~\statbox{0.98,0.99,0.98}{2\%} & -1\%~$\downarrow$\\
	\hline Other & 207~\statbox{0.62,0.81,0.62}{38\%} & +8\%~$\uparrow$\\
	\hline 
	\hline \textbf{Interaction Technique} & \textbf{525}~\statbox{0.03,0.52,0.03}{97\%} & \\ 
	\hline Select & 491~\statbox{0.09,0.55,0.09}{91\%} & +9\%~$\uparrow\uparrow$\\
	\hline Explore / Browse & 445~\statbox{0.18,0.59,0.18}{82\%} & -3\%~$\downarrow$\\
	\hline Reconfigure & 220~\statbox{0.59,0.80,0.59}{41\%} & +4\%~$\uparrow$\\
	\hline Encode & 294~\statbox{0.46,0.73,0.46}{54\%} & -2\%~$\downarrow$\\
	\hline Filter / Query & 314~\statbox{0.42,0.71,0.42}{58\%} & +1\%~$\uparrow$\\
	\hline Abstract / Elaborate & 354~\statbox{0.35,0.67,0.35}{65\%} & -24\%~$\downarrow\downarrow\downarrow$\\
	\hline Connect & 395~\statbox{0.27,0.64,0.27}{73\%} & +9\%~$\uparrow\uparrow$\\
	\hline Guide / Sheperd & 169~\statbox{0.69,0.84,0.69}{31\%} & +7\%~$\uparrow$\\
	\hline Verbalize & 31~\statbox{0.94,0.97,0.94}{6\%} & +1\%~$\uparrow$\\
	\hline 
	\end{tabular} 
	\end{minipage}
	\hspace{1mm}
	\begin{minipage}[t]{0.28\textwidth}
	\begin{tabular}[t]{|l|r|c|}
	\hline \textbf{Visual Variable} & \textbf{542}~\statbox{0.00,0.50,0.00}{100\%} & \\ 
	\hline Color & 540~\statbox{0.00,0.50,0.00}{100\%} & +2\%~$\uparrow$\\
	\hline Opacity & 294~\statbox{0.46,0.73,0.46}{54\%} & +12\%~$\uparrow\uparrow$\\
	\hline Position / Orientation & 83~\statbox{0.85,0.93,0.85}{15\%} & -14\%~$\downarrow\downarrow$\\
	\hline Shape & 89~\statbox{0.84,0.92,0.84}{16\%} & -3\%~$\downarrow$\\
	\hline Size & 164~\statbox{0.70,0.85,0.70}{30\%} & -4\%~$\downarrow$\\
	\hline Texture & 30~\statbox{0.94,0.97,0.94}{6\%} & -3\%~$\downarrow$\\
	\hline 
	\hline \textbf{Evaluation} & \textbf{542}~\statbox{0.00,0.50,0.00}{100\%} & \\ 
	\hline Standard & 132~\statbox{0.76,0.88,0.76}{24\%} & +5\%~$\uparrow$\\
	\hline Comparative & 27~\statbox{0.95,0.98,0.95}{5\%} & -1\%~$\downarrow$\\
	\hline Before / During Development & 140~\statbox{0.74,0.87,0.74}{26\%} & +8\%~$\uparrow$\\
	\hline After Development & 175~\statbox{0.68,0.84,0.68}{32\%} & +8\%~$\uparrow$\\
	\hline Not Evaluated & 196~\statbox{0.64,0.82,0.64}{36\%} & -15\%~$\downarrow\downarrow$\\
	\hline 	
	\hline \textbf{Trust Levels (TL) 1--5} & \textbf{542}~\statbox{0.00,0.50,0.00}{100\%} & \\ 
	\hline Source Reliability & 43~\statbox{0.92,0.96,0.92}{8\%} & +2\%~$\uparrow$\\
	\hline Transparent Collection Process & 21~\statbox{0.96,0.98,0.96}{4\%} & +1\%~$\uparrow$\\
	\noalign{
	\global\dimen1\arrayrulewidth
	\global\arrayrulewidth1pt
	}
	\hline
	\noalign{
	\global\arrayrulewidth\dimen1 
	}Uncertainty Awareness & 84~\statbox{0.85,0.93,0.85}{15\%} & 0\%~~-\\
	\hline Equality / Data Bias & 57~\statbox{0.89,0.95,0.89}{11\%} & +3\%~$\uparrow$\\
	\hline Comparison (of Structures) & 271~\statbox{0.50,0.75,0.50}{50\%} & +7\%~$\uparrow$\\
	\hline Guidance / Recommendations & 142~\statbox{0.74,0.87,0.74}{26\%} & +3\%~$\uparrow$\\
	\hline Outlier Detection & 174~\statbox{0.68,0.84,0.68}{32\%} & -2\%~$\downarrow$\\
	\noalign{
	\global\dimen1\arrayrulewidth
	\global\arrayrulewidth1pt
	}
	\hline
	\noalign{
	\global\arrayrulewidth\dimen1 
	}Familiarity & 24~\statbox{0.96,0.98,0.96}{4\%} & +2\%~$\uparrow$\\
	\hline Understanding / Explanation & 314~\statbox{0.42,0.71,0.42}{58\%} & +10\%~$\uparrow\uparrow$\\
	\hline Debugging / Diagnosis & 129~\statbox{0.76,0.88,0.76}{24\%} & -3\%~$\downarrow$\\
	\hline Refinement / Steering & 164~\statbox{0.70,0.85,0.70}{30\%} & -5\%~$\downarrow$\\
	\hline Comparison & 114~\statbox{0.79,0.89,0.79}{21\%} & -10\%~$\downarrow\downarrow$\\
	\hline Knowledgeability & 47~\statbox{0.91,0.96,0.91}{9\%} & +4\%~$\uparrow$\\
	\hline Fairness & 29~\statbox{0.95,0.97,0.95}{5\%} & +2\%~$\uparrow$\\
	\noalign{
	\global\dimen1\arrayrulewidth
	\global\arrayrulewidth1pt
	}
	\hline
	\noalign{
	\global\arrayrulewidth\dimen1 
	}Experience & 31~\statbox{0.94,0.97,0.94}{6\%} & +2\%~$\uparrow$\\
	\hline In Situ Comparison & 135~\statbox{0.75,0.87,0.75}{25\%} & -2\%~$\downarrow$\\
	\hline Performance & 338~\statbox{0.38,0.69,0.38}{62\%} & +8\%~$\uparrow$\\
	\hline What-if Hypotheses & 87~\statbox{0.84,0.92,0.84}{16\%} & -4\%~$\downarrow$\\
	\hline Model Bias & 58~\statbox{0.89,0.95,0.89}{11\%} & -1\%~$\downarrow$\\
	\hline Model Variance & 34~\statbox{0.94,0.97,0.94}{6\%} & -2\%~$\downarrow$\\
	\noalign{
	\global\dimen1\arrayrulewidth
	\global\arrayrulewidth1pt
	}
	\hline
	\noalign{
	\global\arrayrulewidth\dimen1 
	}Agreement of Colleagues & 21~\statbox{0.96,0.98,0.96}{4\%} & -1\%~$\downarrow$\\
	\hline Visualization Evaluation & 301~\statbox{0.44,0.72,0.44}{56\%} & +12\%~$\uparrow\uparrow$\\
	\hline Metrics Validation / Results & 413~\statbox{0.24,0.62,0.24}{76\%} & +11\%~$\uparrow\uparrow$\\
	\hline User Bias & 24~\statbox{0.96,0.98,0.96}{4\%} & 0\%~~-\\
	\hline 
	\hline \textbf{Target Group} & \textbf{542}~\statbox{0.00,0.50,0.00}{100\%} & \\ 
	\hline Beginners & 104~\statbox{0.82,0.91,0.82}{19\%} & -2\%~$\downarrow$\\
	\hline Practitioners / Domain Experts & 424~\statbox{0.22,0.61,0.22}{78\%} & -3\%~$\downarrow$\\
	\hline Developers & 113~\statbox{0.79,0.90,0.79}{21\%} & +3\%~$\uparrow$\\
	\hline ML Experts & 204~\statbox{0.62,0.81,0.62}{38\%} & +1\%~$\uparrow$\\
	\hline 	
	\end{tabular} 
 	\bigbreak
	Symbol Legend: 
    \\ \thinspace
        +/-[1 -- 8]\% $\uparrow$/$\downarrow$ \thinspace
        \\ +/-[9 -- 16]\% $\uparrow\uparrow$/$\downarrow\downarrow$ \thinspace
        \\ +/-[17 -- 24]\% $\uparrow\uparrow\uparrow$/$\downarrow\downarrow\downarrow$
	\end{minipage} 

	\medskip
	\scriptsize \emph{Note:} Each row shows the total count of techniques per category (with heatmap-style icons), and the \% difference compared to the 2020 STAR.
	\label{tab:categorization-summary}
\end{table*}

\section{IN-DEPTH ANALYSIS}

\subsection{Trust in ML Visualization Revisited}
Table~\ref{tab:categorization-summary} shows the most prevalent aspects of existing visualization techniques for increasing trust in ML.

For \emph{Data}, \emph{computer vision} ($\approx$$33\%$ of 542 articles; \textcolor{ForestGreen}{$\uparrow$} compared to the 2020 STAR), \emph{health} ($\approx$$17\%$; \textcolor{ForestGreen}{$\uparrow$}), \emph{business} ($\approx$$16\%$; \textcolor{ForestGreen}{$\uparrow$}), and \emph{humanities} ($\approx$$15\%$; \textcolor{red}{$\downarrow$}) are the most prominent \emph{domains}, while the \emph{biology}'s popularity \textcolor{red}{$\downarrow$}. \emph{Multi-class classification} ($\approx$$55\%$; \textcolor{red}{$\downarrow$}\textcolor{red}{$\downarrow$}) is still by far the most frequently found \emph{target variable}.

For \emph{ML methods}, \emph{non-linear} ($\approx$$17\%$; \textcolor{red}{$\downarrow$}\textcolor{red}{$\downarrow$}) and then \emph{linear DR} ($\approx$$15\%$; \textcolor{red}{$\downarrow$}\textcolor{red}{$\downarrow$}) techniques are commonly used (the opposite was true in the 2020 STAR), followed by \emph{bagging (ensemble learning)} ($\approx$$14\%$) and \emph{CNNs} ($\approx$$13\%$) from the \emph{DL} subcategory ($\approx$$64\%$; \textcolor{red}{$\downarrow$}\textcolor{red}{$\downarrow$}). The vast majority of the techniques address \emph{supervised learning} ($\approx$$69\%$) and specifically \emph{classification} problems ($\approx$$53\%$; \textcolor{red}{$\downarrow$}), and then \emph{unsupervised learning} ($\approx$$35\%$; \textcolor{red}{$\downarrow$}\textcolor{red}{$\downarrow$}) with \emph{DR} ($\approx$$19\%$; \textcolor{red}{$\downarrow$}\textcolor{red}{$\downarrow$}\textcolor{red}{$\downarrow$}) and \emph{clustering} ($\approx$$13\%$; \textcolor{red}{$\downarrow$}).

The ratio of \emph{in-processing} ($\approx$$29\%$; \textcolor{ForestGreen}{$\uparrow$}) to \emph{post-processing} ($\approx$$67\%$; \textcolor{red}{$\downarrow$}\textcolor{red}{$\downarrow$}) techniques has increased \textcolor{ForestGreen}{$\uparrow$}\textcolor{ForestGreen}{$\uparrow$} from $\approx$$28\%$ to $\approx$$43\%$ in the past three years. Most techniques are built as \emph{model-agnostic} ($\approx$$73\%$; \textcolor{ForestGreen}{$\uparrow$}) to target various ML methods and a bigger user audience.

The absolute majority of the visualizations rely solely on \emph{2D representations} ($\approx$$99\%$; \textcolor{ForestGreen}{$\uparrow$}). For \emph{visual aspects} and \emph{granularity}, almost all techniques used have at least a \emph{computed} ($\approx$$92\%$; \textcolor{red}{$\downarrow$}) component that is not directly \emph{mapped} data ($\approx$$50\%$; \textcolor{red}{$\downarrow$}). \emph{Aggregated information} ($\approx$$86\%$; \textcolor{red}{$\downarrow$}) is more common than \emph{instance-based/individual} instances' exploration ($\approx$$70\%$; \textcolor{red}{$\downarrow$}).

Popular visualizations are \emph{scatterplots} ($\approx$$54\%$; \textcolor{red}{$\downarrow$}), \emph{bar charts} ($\approx$$45\%$; \textcolor{ForestGreen}{$\uparrow$}), \emph{histograms} ($\approx$$32\%$; \textcolor{ForestGreen}{$\uparrow$}), \emph{line charts} ($\approx$$28\%$), \emph{heatmaps} ($\approx$$27\%$; \textcolor{ForestGreen}{$\uparrow$}), \emph{gylphs/icons} ($\approx$$24\%$; \textcolor{red}{$\downarrow$}), and \emph{node-link diagrams} ($\approx$$24\%$). Simpler ones, e.g., \emph{tables/lists} ($\approx$$37\%$; \textcolor{red}{$\downarrow$}) and \emph{matrices} ($\approx$$28\%$; \textcolor{ForestGreen}{$\uparrow$}), work better with \emph{instance-based} exploration.

On the interaction side, \emph{selection} ($\approx$$91\%$; \textcolor{ForestGreen}{$\uparrow$}\textcolor{ForestGreen}{$\uparrow$}), \emph{exploration} ($\approx$$82\%$; \textcolor{red}{$\downarrow$}), and \emph{connection} ($\approx$$73\%$; \textcolor{ForestGreen}{$\uparrow$}\textcolor{ForestGreen}{$\uparrow$}) between different views are the three most common categories found in numerous articles, followed by other interaction techniques, such as \emph{abstraction/elaboration} ($\approx$$65\%$; \textcolor{red}{$\downarrow$}\textcolor{red}{$\downarrow$}\textcolor{red}{$\downarrow$}), \emph{filtering out/searching for} ($\approx$$58\%$; \textcolor{ForestGreen}{$\uparrow$}) specific instances, and \emph{encoding} ($\approx$$54\%$; \textcolor{red}{$\downarrow$}).

\emph{Color} ($\approx$$100\%$; \textcolor{ForestGreen}{$\uparrow$}) is the primary visual channel employed for conveying information in various VA tools and systems. The extensive utilization of \emph{opacity} ($\approx$$54\%$; \textcolor{ForestGreen}{$\uparrow$}\textcolor{ForestGreen}{$\uparrow$}) for hiding data points/instances as well as \emph{size/area} ($\approx$$30\%$; \textcolor{red}{$\downarrow$}) for representing data variables can be attributed to the widespread use of \emph{scatterplots}.

As for the \emph{evaluation}, $\approx$$36\%$ of the visualization techniques we analyzed have \emph{not been evaluated by users} (\textcolor{red}{$\downarrow$}\textcolor{red}{$\downarrow$}), which is a huge improvement from the $51\%$ three years before. If we consider that the unevaluated visualization techniques could undergo extensive quantitative testing with different (synthetic) benchmarking datasets (e.g., a common approach for DR visualization), this shows a huge shift in how important the visualization community considers evaluation.

For TL1, more works tackle \emph{source reliability} problems ($\approx$$8\%$; \textcolor{ForestGreen}{$\uparrow$}) rather than the \emph{transparent collection process} challenge ($\approx$$4\%$; \textcolor{ForestGreen}{$\uparrow$}). For TL2, researchers focus on the \emph{comparison of structures} ($50\%$; \textcolor{ForestGreen}{$\uparrow$}), an increase of $7\%$ prior to the past three years, and then \emph{outlier detection} ($\approx$$32\%$; \textcolor{red}{$\downarrow$}) and \emph{guidance/recommendations} ($\approx$$26\%$; \textcolor{ForestGreen}{$\uparrow$}). For TL3, \emph{understanding} ($\approx$$58\%$; \textcolor{ForestGreen}{$\uparrow$}\textcolor{ForestGreen}{$\uparrow$}), \emph{steering} ($\approx$$30\%$; \textcolor{ForestGreen}{$\uparrow$}), \emph{debugging} ($\approx$$24\%$; \textcolor{red}{$\downarrow$}), and \emph{comparing} ($\approx$$21\%$; \textcolor{red}{$\downarrow$}\textcolor{red}{$\downarrow$}) ML methods are quite popular. \emph{Understanding} became $\approx$$10\%$ more common recently, and more techniques are designed for \emph{debugging} than \emph{comparing} in the past three years. For TL4, \emph{performance} ($\approx$$62\%$; \textcolor{ForestGreen}{$\uparrow$}), \emph{in situ comparison} ($\approx$$25\%$; \textcolor{red}{$\downarrow$}), and \emph{what-if hypotheses} ($\approx$$16\%$; \textcolor{red}{$\downarrow$}) are commonly emerging categories associated with a concrete ML model selection. For TL5, \emph{metrics validation and results observation} is the most frequent category covering $\approx$$76\%$ of all the articles (\textcolor{ForestGreen}{$\uparrow$}\textcolor{ForestGreen}{$\uparrow$}).

The visualization techniques have as a main \emph{target group}, usually \emph{practitioners/domain experts} ($\approx$$78\%$; \textcolor{red}{$\downarrow$}), followed by \emph{ML experts} ($\approx$$38\%$; \textcolor{ForestGreen}{$\uparrow$}) with a huge difference. The most underrepresented groups are model developers ($\approx$$21\%$; \textcolor{ForestGreen}{$\uparrow$}) and beginners ($\approx$$19\%$; \textcolor{red}{$\downarrow$}).

In summary, the ML side primarily relies on \emph{model performance} and \emph{metrics validation} to increase trust in ML, and the visualization side often opts for \emph{scalable multivariate visualizations} preferred by experts.

\begin{figure*}
\centerline{\includegraphics[width=36pc]{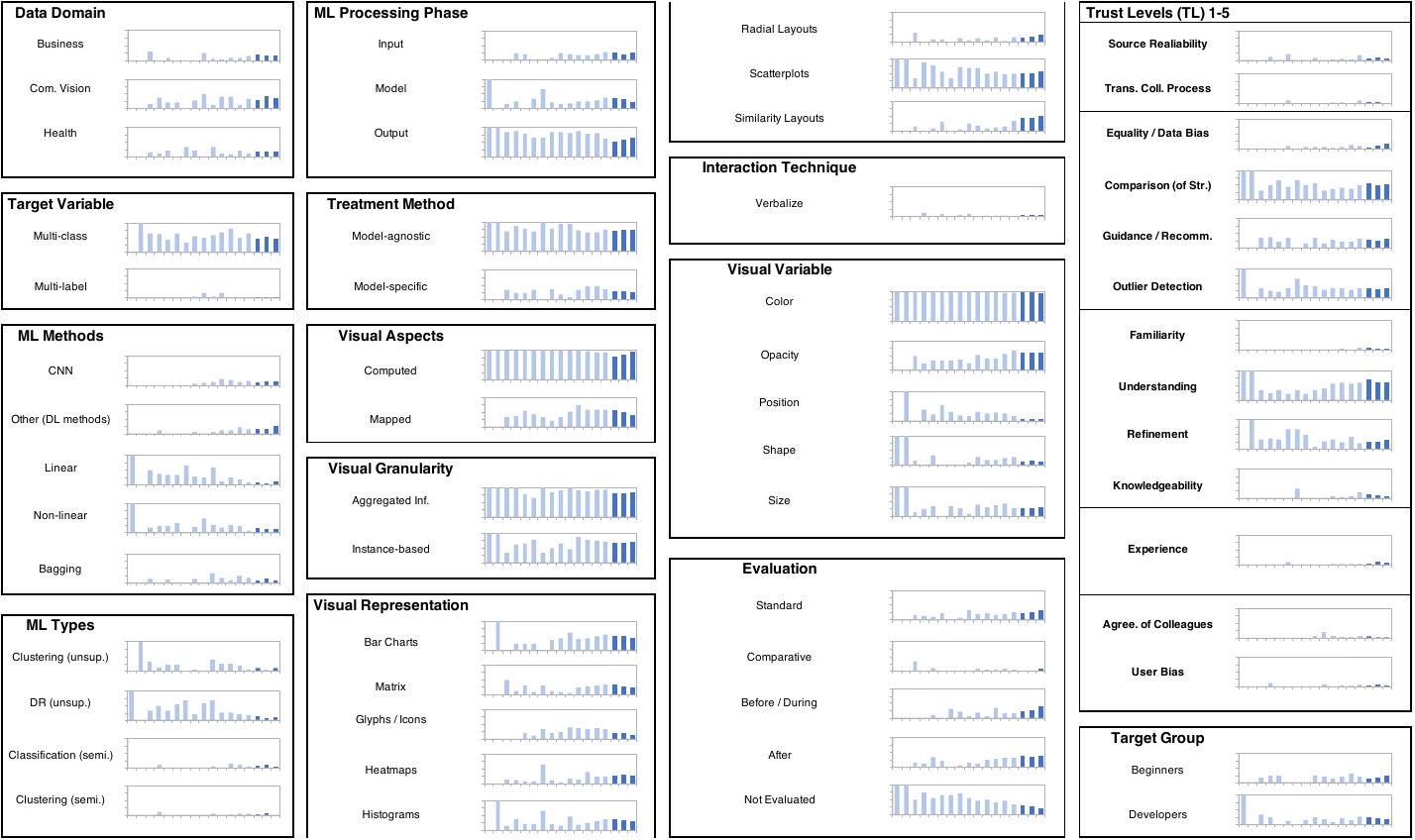}} \vspace{-1mm}
\caption{Sparklines visualizing the relative category popularity from 2007 to 2023 for the categories discussed in this article. The complete plot can be found in the supplemental material. Each bar shows the support for each category within our dataset (see Table~\ref{tab:categorization-summary}) in relation to the total count of techniques for the same year (in light blue; the last three years are in dark blue).}
\label{fig:temporal} \vspace{-1mm}
\end{figure*}

\subsection{Temporal Trends}
We have also analyzed the temporal distribution for each category (normalized per respective year) to find which categories have been more or less prominent in the field in the last three years, as displayed in Figure~\ref{fig:temporal}.

These results suggest that \emph{business}, \emph{computer vision}, and \emph{health} data are the most commonly \emph{targeted domains}. \emph{Multi-class data} is steadily at the forefront of research while \emph{multi-label data} is still an underrepresented category with no trend for a potential increase.

For \emph{ML methods}, we see an increase in \emph{other DL methods}, such as explaining (vision) transformer NNs, as well as CNNs as the specific NN architecture most prominent compared to all others. Techniques for \emph{non-linear DR} are more likely to be researched in recent years compared to \emph{linear DR} methods. \emph{Bagging} is the most common \emph{ensemble learning} method. For \emph{ML types}, we observe a small decline in visualization techniques for \emph{clustering} and \emph{DR}, and a subtle increase in \emph{semi-supervised learning}. A more interesting result here can be observed in the \emph{ML processing phase}, where \emph{in-processing} is visualized increasingly most often, then \emph{pre-processing}, and finally \emph{post-processing}. \emph{Model-agnostic} techniques are increasingly more common in the community compared to \emph{model-specific}. \emph{Computed} and \emph{aggregated information} are gradually rising compared to their counterparts.

For the \emph{visual representation}, the \emph{scatterplots} and \emph{similarity layouts}, \emph{heatmaps}, and \emph{radial layouts} are more and more commonly used compared to \emph{bar charts}, \emph{matrix representations}, \emph{glyph/icons}, and \emph{histograms} that remain constant or slightly decrease in popularity. For the \emph{interaction techniques}, \emph{verbalization} emerged in 2010 and has not drawn much attention in the visualization community---but recently, there has been some re-appearance compared to the past. Furthermore, \emph{opacity}, \emph{shape}, and \emph{size} are the most usual ways to represent the data after \emph{color}, while \emph{position} is almost extinct. Although \emph{evaluating visualizations} with \emph{comparative evaluations} is very rare since each technique does not have a suitable direct counterpart, the number of evaluated techniques increases in total compared to the past---expert interviews before, during, or after tool development, but also user studies.

For the TLs, many techniques are not covered much by articles, such as \emph{transparent collection processes} and \emph{source reliability}. Notably, \emph{equality/data bias} started to emerge as a theme in the few articles published in 2023. The visualization community could also explore other intriguing research topics, such as investigating how visualization can assist with the \emph{familiarity} a user has for a \emph{learning method}. \emph{Comparison of structures}, \emph{guidance}, and \emph{outlier detection} seem to be in the spotlight based on Figure~\ref{fig:temporal}. Additionally, \emph{understanding} and \emph{refinement} of \emph{learning methods} appear to peak in the last year. \emph{Knowledgeability} about \emph{learning methods} and details available to diverse user types are slightly better supported than in the past. Moreover, customizable and reconfigurable visualizations that consider the users' \emph{experience} level to select a specific ML model remain largely underresearched. A few visualization techniques enable \emph{agreement of colleagues} and evaluate the use of provenance in VA tools and systems to cover trust in ML issues. \emph{User bias} is absent from almost all VA systems.

Finally, \emph{beginners} and \emph{model developers} are the two least represented \emph{target groups}.

\subsection{Category Correlations \& Patterns}
We have conducted a correlation analysis for our categorization (see Table~\ref{tab:categorization-summary}). Individual articles were the instances/observations, and categories were the dimensions/variables. We employed linear correlation analysis to gauge the relationship between category pairs. Given the extensive size of the correlation matrix, we have only provided a small thumbnail representation of it in Table~\ref{tab:correlation}(a), with the complete table available in the supplemental material. Table~\ref{tab:correlation}(b) shows Pearson's \emph{r} coefficient values and highlights distinctive patterns, as well as notable instances of positive (green) and negative (red) correlations among the categories.

In the remainder of this subsection, we focus on the correlations different from our previous survey article published in 2020,$^9$ but for the sake of completeness, we refer the reader to Table~\ref{tab:correlation} so as to check all interesting correlations found in our dataset's entirety.

The new findings for \emph{negative} correlation is that techniques \emph{not evaluated by users} do not involve \emph{visualization interactions}, supporting our previous claim that static visualizations are being evaluated via quantitative experiments (if at all). For example, the correlation between \emph{linear} and \emph{non-linear DR} with no evaluation is because of the use of synthetic datasets for quantitative experiments. Additionally, \emph{beginners} rarely use \emph{abstract} as an interaction technique.

\begin{table}
\caption{The correlation matrix for the categories in Table~\ref{tab:categorization-summary}. (a) shows all categories' pair-wise correlations as a thumbnail. (b) shows important findings, with cases sorted based on absolute correlation strengths and opposed to our 2020 STAR.}
\label{tab:correlation}
\centerline{\includegraphics[width=18.5pc]{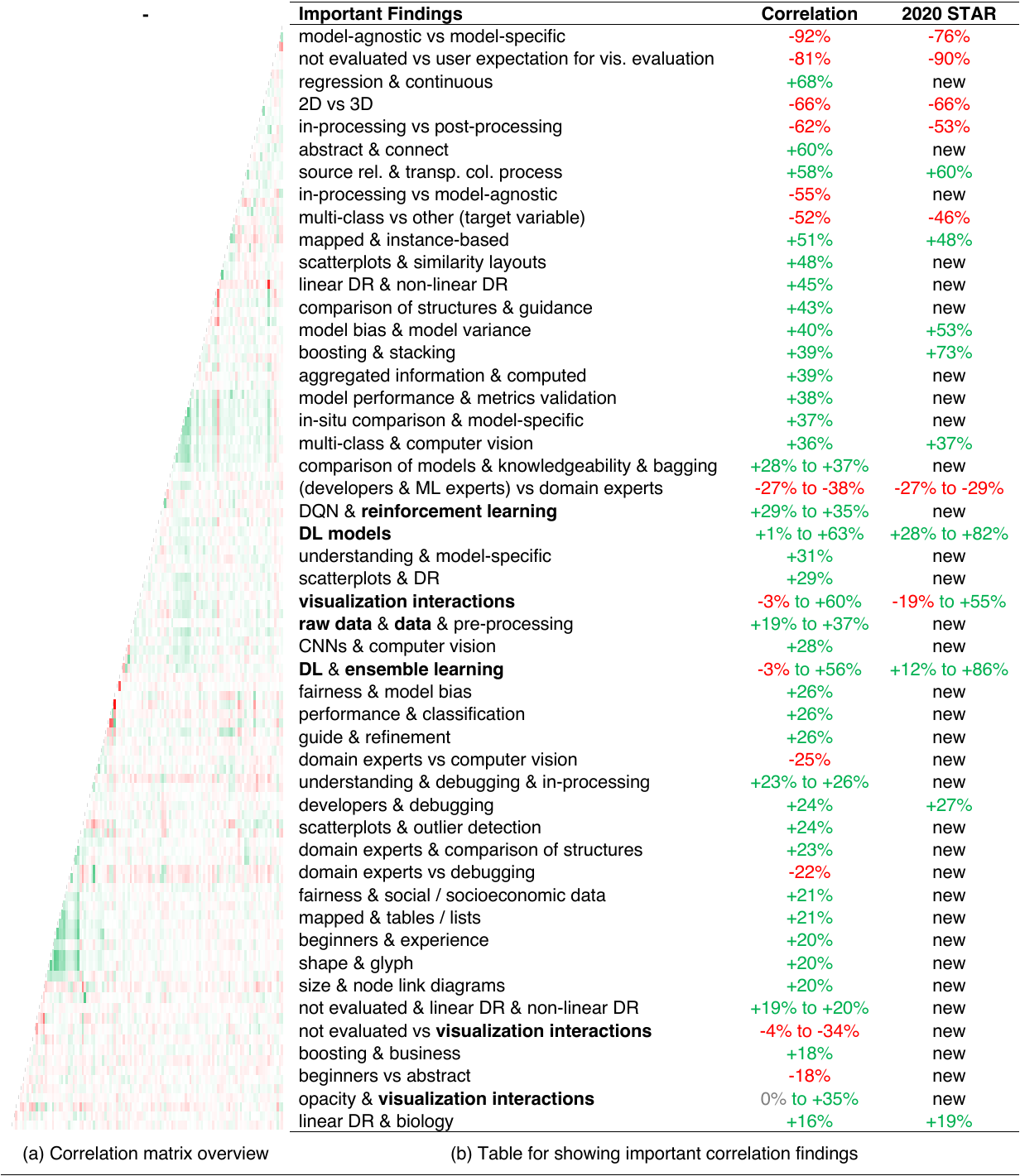}}
\medskip
\scriptsize \emph{Note:} We use red for negative correlations and green for positive ones. Cases highlighted in \textbf{bold} refer to broader groups of categories.
\end{table}

Cases with \emph{positive} correlation start with an obvious case of a \emph{continuous target variable} being associated with \emph{regression} problems. 
\emph{Abstracting} further the data leads to more in-depth elaboration with different views that require to be better \emph{connected}.

\emph{Similarity layouts} are usually represented with \emph{scatterplots} compared to other visualizations (e.g., a similarity-based matrix). 
The \emph{comparison of structures} is a moderately correlated category with the goal to provide further \emph{guidance} at the data exploration phase.

\begin{table*}
\caption{The top eight terms and their weights for each of the respective ten topics created by LDA when applied to all articles.}
\label{tab:topics}
\centerline{\includegraphics[width=37pc]{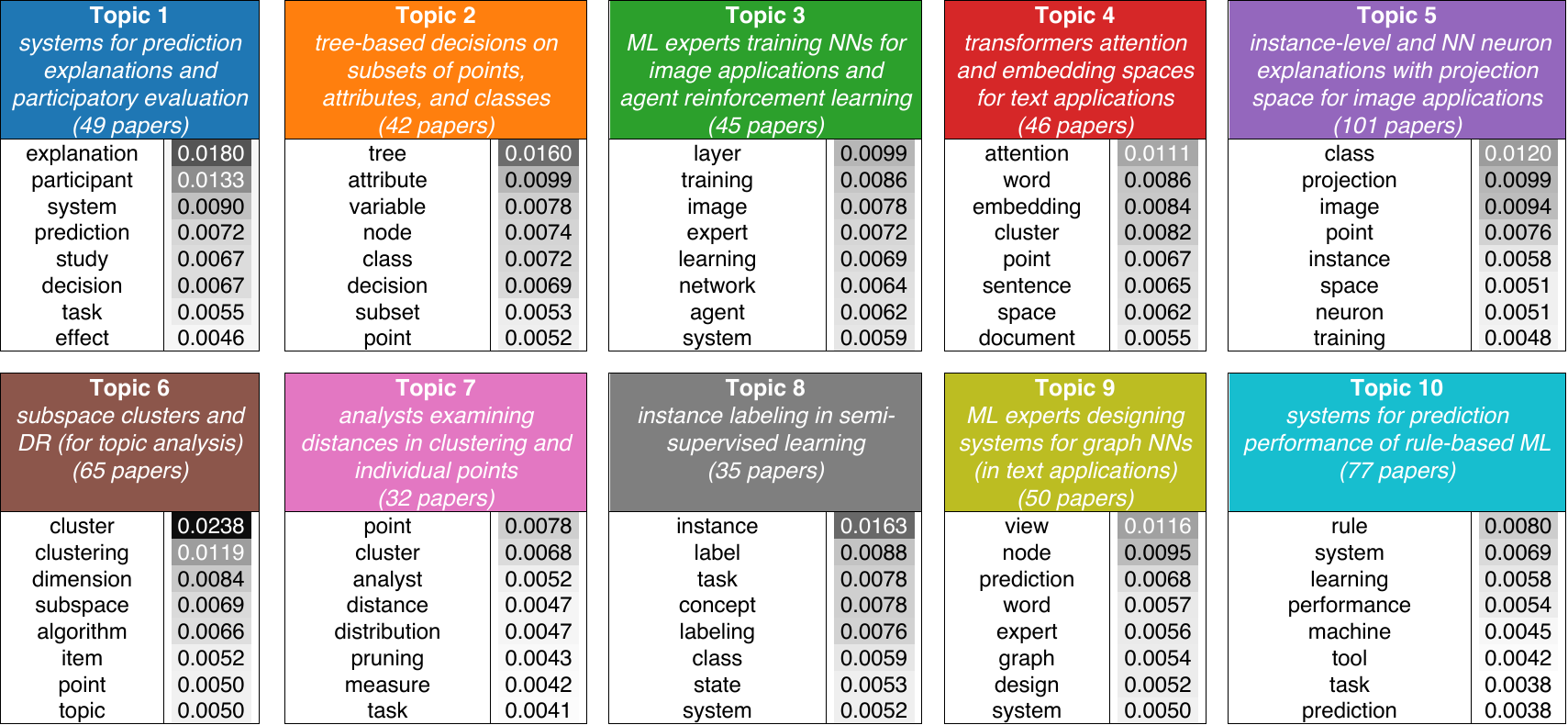}}
\smallskip \scriptsize \emph{Note:} The proposed topic titles are in italics. Every topic is encoded by one distinct color. A gray colormap is used for the weight of each term.
\end{table*}

\emph{Computed} data and statistics lead to \emph{aggregated information} visualization. 
The \emph{model performance} is typically tested with \emph{metrics validation} to choose a concrete ML model.
\emph{In situ comparison of models} occurs most probably for examining \emph{model-specific} ML algorithms.
The \emph{comparison of models} correlates with \emph{knowledgeability} and \emph{bagging} since knowing about how different weak learners work is necessary to use them in an ensemble.
\emph{Understanding} an ML model weakly correlates with techniques for \emph{specific models}.

Visualizing 2D \emph{DR} projections with \emph{scatterplots} is rather usual. Additionally, visualization techniques for \emph{pre-processing} are correlated with the TLs of \emph{raw data} and \emph{data} as a whole. 
\emph{CNNs} are the most common ML algorithms for \emph{computer vision}. 
Investigating \emph{fairness} and \emph{model bias} in ML reassures that models and systems operate equitably without replicating existing biases or introducing new ones, thus making more robust predictions.
\emph{Model performance} in \emph{classification} problems aims to monitor (and increase) the model accuracy.
\emph{Guide} and \emph{refinement} visualization loops are important for well-calibrating ML models.

Gaining a deeper \emph{understanding} and effectively \emph{debugging in-processing} mechanisms of ML models are central to improving their reliability and usability. 
\emph{Scatterplots} are invaluable for \emph{outlier detection}, offering visual insights into data distributions and aiding in identifying anomalies that might otherwise go unnoticed.
\emph{Domain experts} are uniquely positioned to \emph{compare} and analyze various \emph{structural} frameworks, leveraging their specialized knowledge to offer deep insights and critiques. Addressing \emph{fairness} in the context of \emph{social and socioeconomic data} tries to ensure that ML technologies do not worsen societal inequalities and disparities.
\emph{Mapped tables and lists} contribute to a more structured and organized representation of data, enhancing accessibility and comprehension for diverse users. 
For \emph{beginners}, accumulating hands-on \emph{experience} is essential in solidifying their understanding and promoting skill development.
We also found a weak correlation of \emph{shape} to design \emph{glyphs} in visualizations
\emph{Size} is crucial in \emph{node-link diagrams}, where the dimensions can effectively convey hierarchical relationships and relative importance among interconnected entities.



Furthermore, while the correlation analysis results discussed above focus on pairs of categories, we also conducted frequent pattern mining with our survey data to uncover interactions between multiple categories. 
Focusing on the patterns with at least 10\% support in the data (as the number of possible category combination patterns is otherwise overwhelming), we have detected 12 patterns with 15 categories recurring in our data (the list is provided as part of supplemental materials), while further patterns with 14 or fewer categories could also be considered. 
The top pattern in the current survey data is supported by 66 articles (12\%), and it provides a glimpse into the profile of typical VA approaches in this field: 
\emph{2D} visual representations for \emph{computed} aspects and \emph{aggregated information}; support for \emph{select}, \emph{explore/browse}, \emph{encode}, \emph{filter/query}, \emph{abstract/elaborate}, and \emph{connect} tasks; use of \emph{color} and \emph{opacity} visual channels; involving the \emph{performance} aspects, \emph{visualization evaluation} as well as \emph{metrics validation/results}, and \emph{practitioners/domain experts}.

\subsection{Topic Modeling}
Following the same methodology as in our previous articles,$^8$$^,$$^9$ we have conducted a topic analysis using latent Dirichlet allocation (LDA) based on the full texts from the PDFs. This approach assigns documents to different clusters with various weights (see supplemental material) and identifies descriptive terms for each cluster. While the topic modeling outcomes are partly influenced by the chosen parameters, they offer insights that complement those obtained from our manual analysis. Thus, the topic analysis confirms and deepens our understanding of the categorized articles.

\begin{figure*}
\centerline{\includegraphics[width=37pc]{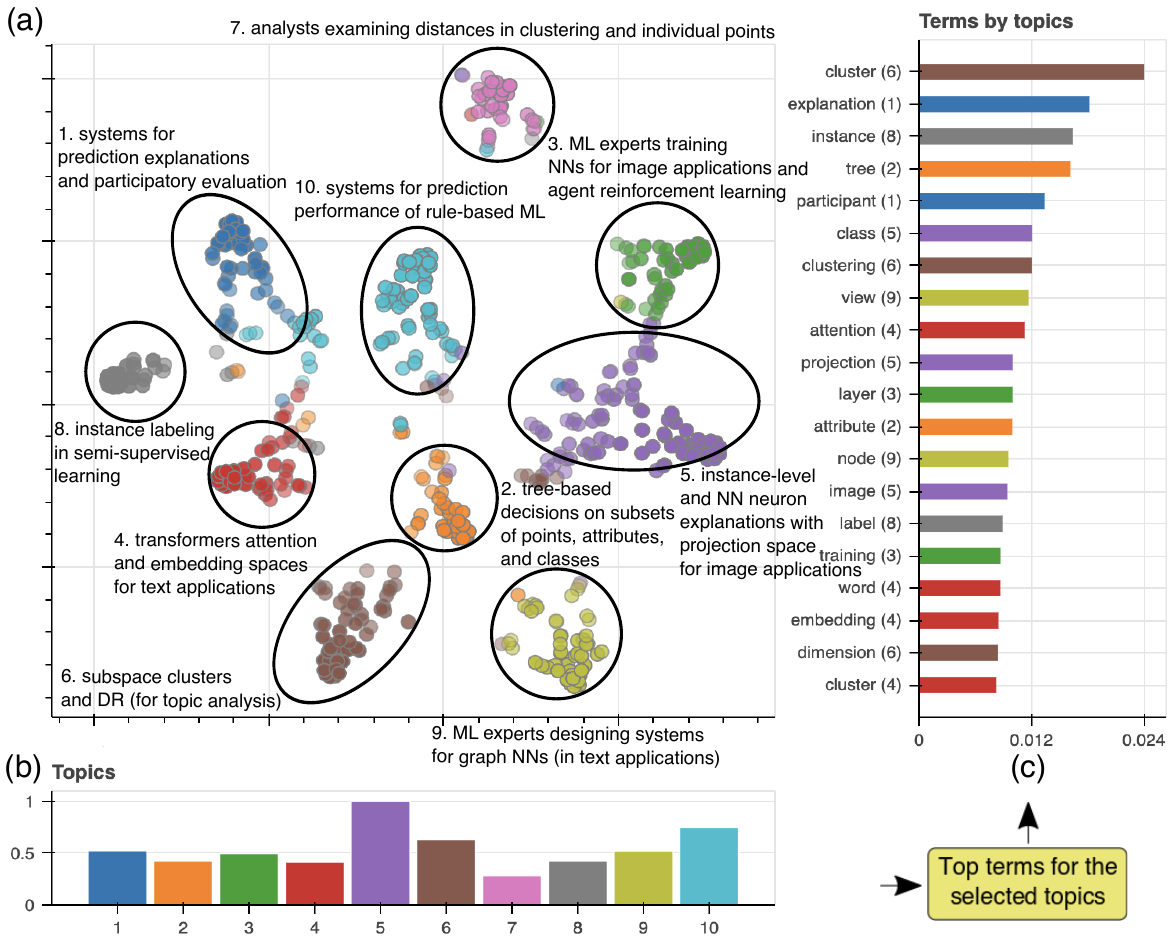}} \vspace{-3mm}
\caption{Visualization of the ten topics extracted from the 542 articles. (a) t-SNE projection of articles, color-coded according to their topics. The black outlines were manually added over the image, and the tags are the full topic titles. (b) Bar chart of the summed-up weights for all articles per topic (scaled between 0 and 1). (c) Horizontal bar chart of the top terms across all topics. Here, topics are double-encoded with color and number (in parentheses); a single term might appear in multiple topics.}
\label{fig:topics-projection} \vspace{-1mm}
\end{figure*}

In our approach, documents get assigned into \emph{one of ten clusters} according to the topmost weight, with eight descriptive terms per cluster, as shown in Table~\ref{tab:topics}. Based on a thorough review of the top terms and the content of the articles, we manually assigned the \emph{topic titles} for each topic. Finally, the results are visualized as a DR projection and bar charts (see Figure~\ref{fig:topics-projection}).$^8$$^,$$^9$

\subsection{Topics}
Here, we summarize the ten topics (see Table~\ref{tab:topics}).

\textcolor{ColT1}{\rule{\boxh}{\boxh}} T1
This topic comprises 49 articles, focusing on the \emph{understanding/explanation} of ML models. Multiple visualization tools belonging to this category have been evaluated with user studies involving diverse participants, such as domain and ML experts. Additionally, an unresolved challenge related to this topic is determining the most effective strategies for developing visualization tools that account for ML trustworthiness.

\textcolor{ColT2}{\rule{\boxh}{\boxh}} T2
A common theme here is the use of visualization in explaining decision trees (found in 42 articles). Other subtopics are the exploration of behavior regarding the decomposition of instances (or points if DR is used), the user's role in reasoning with subsets of attributes, and showing how classes are formed in connection to the internal parts of rule-based models.

\textcolor{ColT3}{\rule{\boxh}{\boxh}} T3
The 45 articles focus mainly on DL for image data. A recent subtopic here is federated learning, that is, a decentralized training approach allowing distant clients to learn from data while private information is kept hidden. Visualization can help inspect and detect anomalies and errors during training ML algorithms in federated learning. Moreover, this topic is connected to reinforcement learning problems, where more research is needed to study what behaviors are associated with low and high reward levels and to track their evolution throughout the training phase.

\textcolor{ColT4}{\rule{\boxh}{\boxh}} T4 
This topic has 46 articles allocated, focusing on the use of projections for explaining NN architectures. For example, the recent trend of using transformers for text data---and even extending their use in image applications with the so-called vision transformers---is covered by this topic, with new VA solutions being invented to check how they work.

\textcolor{ColT5}{\rule{\boxh}{\boxh}} T5 
In DL, significant research efforts have been devoted to understanding the activation of neurons within NNs and visually representing them. Various visualizations, such as 2D saliency and activation maps, have been widely utilized for depicting the activation levels of neurons in diverse DL models, particularly those related to image processing. This topic contains 101 articles that discuss techniques for visualizing the inner workings of NNs during training to detect how they behave in specific instances using projections.

\textcolor{ColT6}{\rule{\boxh}{\boxh}} T6 
The projection of points with DR is useful for guiding subspace exploration. VA systems are developed for analyzing projection segments to perform error analysis and mine insightful patterns from it. This subtopic and the other about visual analysis of relations between points and dimensions within the various projection spaces were found in 65 articles.

\textcolor{ColT7}{\rule{\boxh}{\boxh}} T7
Another area of exploration with 32 articles involves identifying the appropriate distance function. It is essential to ensure that these distances are maintained in the 2D projection generated from the high-dimensional space. This aspect should align with the \emph{users' cognitive expectations} of observable clusters.

\textcolor{ColT8}{\rule{\boxh}{\boxh}} T8
Incremental data labeling and active learning are a few common themes in this topic of 35 articles. VA systems can suggest which unlabeled instances should be picked first and why to achieve improved predictive performance. Counterfactual explanations are also important for users with systems proposing specific examples that are worthy of investigation.

\textcolor{ColT9}{\rule{\boxh}{\boxh}} T9
A prevalent shared aspect among most of the 50 articles in this topic class is their graph NN algorithms and RNNs for text data. 
Delving into the hidden states of these networks appears to recover lost information that could increase ML trustworthiness under the appropriate guidance of an expert. A hidden subtopic with a few techniques is relevant for the \emph{outlier detection} catergory that stands out in our categorization with 174 articles in total, as shown in Table~\ref{tab:categorization-summary}.

\textcolor{ColT10}{\rule{\boxh}{\boxh}} T10
This topic comprises 77 articles that specifically refer to rule-based ML, such as bagging and boosting methods. Rule-based ML models are successful with tabular data and arguably easier for analysts to understand. However, they sometimes reach a level of complexity similar to that of DL, thus it can still be difficult to simplify them.

Compared to our 2020 STAR,$^9$ Topics 2, 4, 8, 9, \&~10 are entirely new due to the recent advancements in graph NNs and (vision) transformers, as well as the need for simpler tree- and rule-based systems for transparent decision-making and ML explanation. The remaining topics are similar but slightly changed compared to the 2020 counterparts. Notably, Topic 1 now refers explicitly to explaining the predictions rather than the ML's inner workings and evaluating with user studies. Topic 3 is more prominent than before, but hyperparameter tuning is currently disregarded. Explaining individual instances and using DR projections for NN explanation also greatly surged lately (Topic 5).

\subsection{Topic Embedding}
Figure~\ref{fig:topics-projection}(a) presents a 2D projection of the ten-dimensional space of topics. Figure~\ref{fig:topics-projection}(b) reveals that Topics 5~\&~10 cover $\approx$$33\%$ of all articles, followed by Topics 6~\&~1, 9 \&~3, and the others. Based on Figure~\ref{fig:topics-projection}(c), some interesting top terms are ``clusters'', ``explanations'' (cf. \emph{understanding} in Table~\ref{tab:categorization-summary}), ``instances'' (\emph{instance-based visualization}), ``trees'' (\emph{bagging}), ``participants'' (\emph{evaluation}), ``projections'' (\emph{DR}), ``layers'' (\emph{DL}), and ``image data'' (\emph{computer vision}).

The t-SNE projection in Figure~\ref{fig:topics-projection}(a) shows the purest clusters color-encoded in brown and gray, which are about unique topics (Topics 6~\&~8, respectively). Additionally, the misclassification of blue (Topic 1) \& cyan (Topic 10) points along with green (Topic 3) \& purple (Topic 5) points in the projection occurs because of two conceptual terms shared across these topic pairs, namely, the terms ``systems'', ``prediction'' for 1 \& 10 and ``NN'' and ``image'' for 3~\&~5. Despite Topic 7 being a specific topic, it is associated with the general term: ``point''. Thus, a few points in the projection are from Topics 2, 5, 8 \&~10 because the individual ``points'' are related to specific ``instances'' (or similar terms).

The automatically produced topics further support our categorization. For instance, Topics 1~\&~10 represent VA systems focusing on the visualization of prediction explanations with participatory evaluation and prediction performance for rule-based ML, respectively, to facilitate \emph{understanding/explanation}. Topic 3 is about \emph{diagnosing/debugging} the reinforcement learning's training procedure. In contrast, Topics 6~\&~7 reflect the \emph{comparison of data structures} using projections/DR and clustering, respectively. Finally, Topic 2 focuses on the \emph{comparison} of different tree-based algorithms.

\section{OPEN CHALLENGES}

To familiarize readers with each topic, we provide representative references per open challenge (O1--O8).

\subsection{O1: Improving Popular XAI Methods}
Most of the well-known XAI approaches that involve the use of visualization for communicating results were made by experts from other fields. The increasing popularity of these model-agnostic techniques lies in their view of a model as a mathematical function. These functions can be simplified, and interpretation methods can then assess these simpler parts' behavior. Visualization researchers can contribute to developing VA systems for improving these methods' interactivity and visual representations,$^1$$^6$ or directly upgrading the techniques' algorithmic components with their expertise.$^1$$^7$

\subsection{O2: New NN Approaches \& Self-Supervision}
The ML community continually invents new NN architectures or applies existing ones in different settings, while visualization researchers are slow to adopt and work on them. For example, transformer NNs have been used in computer vision settings with promising results compared to convolutional NNs, but still, the visualization community has hardly ever experimented with helping users explore them visually.$^1$$^8$$^,$$^1$$^9$ Similarly, self-supervision is another area of rapid advancement that requires the visualization community's attention,$^2$$^0$ as well as prompt engineering and model checkers for ``hallucinations'' of large language models (LLMs).$^2$$^1$

\subsection{O3: Multiverse \& Confirmatory VA}
Most VA solutions, as of recently, were arguably focusing on exploratory visual analysis. There is a trend in utilizing visual representations for confirmatory analysis and hypothesis testing, as well as causal reasoning (e.g., counterfactual reasoning, causal learning, and causal inference), which could be enhanced by visualizations.$^2$$^2$ Another question is how can VA systems guide users through an exhaustive multiverse analysis for choosing the optimal model strategies that will help them understand how a complex model works and decide upon a specific action plan?$^2$$^3$

\subsection{O4: Input/Output Uncertainty Quantification}
Although the majority of the VA systems aim to help users understand models, quantifying the uncertainty of input and output, as well as checking their robustness and sensitivity to changes, are two major challenges because people want to deploy dynamical, well-calibrated models in real-world practical scenarios.$^2$$^4$ While uncertainty quantification and sensitivity analysis are popular in other fields, visualization research is still limited in checking the inputs and outputs of models, such as with the conformal prediction method that provides statistical guarantees.$^2$$^5$

\subsection{O5: Rigorous Evaluation \& Benchmarking} 
Effective collaboration with domain experts from diverse fields often faces substantial communication challenges. These arise from differences in terminology, expectations, and areas of expertise, demanding huge efforts to align objectives and solutions. To mitigate the need for costly user evaluations, the exploration of AI-assisted, simulated evaluation methods is very important. The state-of-the-art benchmarking datasets should also become more rigorous since they may contain non-negligible errors.$^2$$^6$
The development of rational agent benchmarks that estimate the need for and benefits of visualization can also assist XAI methods evaluation.$^2$$^7$ 
Therefore, such benchmarks potentially minimize the dependency on extensive human participation, thereby reducing associated costs.

\subsection{O6: Model Deployment \& Visual Channels}
Scaling VA systems to manage numerous instances, features, and algorithms presents significant challenges, particularly in multi-class classification scenarios. Progressive VA combined with incremental learning is one notable solution, but alternative methods for designing less complex, but more robust systems are essential in real-world settings, where covariate and label shifts are common phenomena.$^2$$^8$ The concurrent use of visual channels, such as color and opacity, for other encoding tasks increases the complexity. Many users find it difficult to navigate these advanced tools, especially at initial exposure. Visualization literacy can partially alleviate this problem by educating people. Additionally, complementary to existing tutorials, storytelling can further illustrate tool functionalities, while the integration of sonification and verbalization can boost user understanding and ease of use of VA systems.

\subsection{O7: Advancing the Impact \& Reproducibility}
Many application-specific VA systems are never deployed in the real world, making them susceptible to concept and distribution drifts.$^2$$^9$ This practice highlights the need for further exploration into the out-of-domain/distribution applicability of models, emphasizing the urgency to transform specialized models into more versatile ML solutions.$^3$$^0$ Ensuring the accessibility and usability of VA systems for a broader user base can be achieved through various means,$^3$$^1$ including the integration of visualizations into computational notebooks$^3$$^2$ and offering combined programmatic API and interactive UI solutions that elicit user feedback over long periods of time.$^3$$^3$$^,$$^3$$^4$ Here, the adoption of open-source practices is vital since it not only promotes community involvement and contribution, but also addresses significant challenges related to the replication and reproducibility of ML models used in VA systems.

\subsection{O8: Underexplored Areas}
All underrepresented categories may spark novel research ideas for non-TL categories, implicitly influencing trust in ML. For example, the rareness of visualization tools targeting boosting and stacking ensemble learning methods. Multi-label and regression problems receive less attention than classification. In unsupervised learning, association/pattern mining is less covered by VA tools. Reinforcement learning is also mostly ignored. Finally, the targeted users that require further focus are beginners and people of various experience levels who want to analyze their data in general.

\section{DISCUSSION}

A retrospective reflection on the 2020 STAR's open challenges in relation to Table~\ref{tab:categorization-summary} reveals that the visualization community further researched (a) security vulnerabilities with VA systems for federated learning and adversarial attacks, (b) fairness of the model predictions and counterfactual scenarios, but still without an emphasis on alternative strategic action plans for decision-making (O3), and (c) various forms of biases except for intrinsic visualization and user biases (O5).

Reoccurring challenges in this survey article are O6, O7, and O8, but with further concerns regarding visualization research reproducibility, as well as the deployment of models and VA systems in the wild. Additionally, the demand for approaches fostering visualization literacy and advancing the real-world impact with user-friendly systems that encourage interdisciplinarity collaboration is evident.

Finally, the remaining open challenges (O1, O2, O4) are entirely new and were identified by manually reviewing the content of the categorized articles.

\section{CONCLUSION}
In this survey article, we have provided an overview of the updated analyses of the trust in ML visualization techniques dataset maintained by us and provided via an online survey browser. We categorized and conducted various analyses based on the 542 collected, peer-reviewed articles that present a broad range of visualization techniques to improve the trustworthiness of ML models and their outputs. Through our topic, temporal, correlation, and pattern mining analyses, we sought to uncover hidden connections and patterns, as well as identify emerging topics and trends over time for the categorized articles. Our findings reveal the rising enthusiasm for implementing VA tools and systems to enhance trust in ML across diverse data domains, tasks, and interdisciplinary applications. In the future, we plan to keep the online survey browser up to date by continuing to collect and categorize data, especially since LLMs, computer vision, and self-supervision research have recently been thriving with the hope of achieving artificial general intelligence. \vspace*{-8pt}

\section{ACKNOWLEDGMENTS}
This work was partially supported through the ELLIIT environment for strategic research in Sweden.

\def\refname{REFERENCES}

\vspace*{-8pt}

\begin{IEEEbiography}{Angelos Chatzimparmpas}{\,} is a postdoctoral scholar with the Department of Computer Science, Northwestern University, Evanston, IL, 60208, USA. His research interests include visual analytics, human-computer interaction, machine learning interpretability, human-centered AI, and XAI systems. He received his Ph.D. degree in computer and information science from Linnaeus University, Växjö, Sweden, in 2023. He is the corresponding author of this article. Contact him at angelos.chatzimparmpas@northwestern.edu.
\end{IEEEbiography}

\begin{IEEEbiography}{Kostiantyn Kucher}{\,} is a senior lecturer with the Department of Science and Technology, Linköping University, Norrköping, 60233, Sweden. His research interests include visual text and network analytics, explainable AI, and domain applications of information visualization and visual analytics, especially in digital humanities and information science. He received his Ph.D. degree in computer and information science from Linnaeus University, Växjö, Sweden, in 2019. Contact him at kostiantyn.kucher@liu.se.
\end{IEEEbiography}

\begin{IEEEbiography}{Andreas Kerren}{\,} is a full professor with the Department of Science and Technology, Linköping University, Norrköping, 60233, Sweden, and the Department of Computer Science and Media Technology, Linnaeus University, Växjö, 35195, Sweden. He holds the Chair of Information Visualization at LiU and is head of the ISOVIS group at LNU. He is also an ELLIIT professor supported by the Excellence Center at Link{\"o}ping--Lund in Information Technology and key researcher of the Linnaeus University Centre for Data Intensive Sciences and Applications. His research interests include several areas of information visualization and visual analytics, especially visual network analytics, text visualization, and the use of visual analytics for explainable AI. He received his Ph.D. degree in computer science from Saarland University, Saarbrücken, Germany, in 2002. Contact him at andreas.kerren@liu.se.
\end{IEEEbiography}

\end{document}